# Towards THGEM UV-photon detectors for RICH: on single-photon detection efficiency in Ne/CH$_4$ and Ne/CF$_4$

**C. D. R. Azevedo**[a,b][*], **M. Cortesi**[b], **A.V. Lyashenko**[b], **A. Breskin**[b], **R. Chechik**[b], **J. Miyamoto**[b], **V. Peskov**[b,d], **J. Escada**[c,b], **J.F.C.A. Veloso**[a] and **J.M.F. dos Santos**[c]

[a] *I3N – Departamento de Física, Universidade de Aveiro,*
*Campus Universitário de Santiago 3810-193 Aveiro, Portugal*

[b] *Department of Particle Physics, Weizmann Institute of Science,*
*76100, Rehovot, Israel*

[c] *Departamento de Física, Universidade de Coimbra,*
*P-3004-516, Coimbra, Portugal*

[d] *CERN*
*1206 Geneva, Switzerland*
*E-mail:* cdazevedo@ua.pt

ABSTRACT: The article deals with the detection efficiency of UV-photon detectors consisting of Thick Gas Electron Multipliers (THGEM) coated with CsI photocathode, operated in atmospheric Ne/CH$_4$ and Ne/CF$_4$ mixtures. We report on the photoelectron extraction efficiency from the photocathode into these gas mixtures, and on the photoelectron collection efficiency into the THGEM holes. Full collection efficiency was reached in all gases investigated, in some cases at relatively low multiplication. High total detector gains for UV photons, in excess of 10$^5$, were reached at relatively low operation voltages with a single THGEM element. We discuss the photon detection efficiency in the context of possible application to RICH.

---

[*] Corresponding author.

# Contents

# 1. Introduction

The development of advanced imaging Cherenkov detectors is a subject of intensive R&D; examples are novel UV-photon detectors for RICH in CERN-Super LHC-ALICE [1] and CERN-COMPASS [2]. The requirements imposed by these experiments include high sensitivity to single photons, stable operation under intense ionizing-radiation background and the possibility of covering large detection area at low production cost.

The Thick Gas Electron Multiplier (THGEM) could be a favourable building block and an electrode of choice, including one with a reflective CsI photocathode deposited on its top surface (Figure 1) [3, 4]. It is considered as an option for upgrading RICH detectors [5, 6]. THGEM can be produced over large area by simple printed-circuit board (PCB) techniques (mechanical drilling of sub-mm diameter holes, ~1mm apart, in a ~0.5mm thick PCB substrate – followed by chemical etching of the holes rim [7]). The reader is referred to a concise recent review of THGEM detectors with a bibliography to previous works [8]. Furthermore, THGEM-detectors were shown to have moderate (sub-millimeter) localization resolution [9, 10] and about 10ns time resolution [11], which comply with the requirements from most RICH devices in Particle Physics. High gains were demonstrated in a variety of gases including noble gases and their mixtures (Ar, Xe, Ar-Xe [12], $Ar/CH_4$, $Ar/CO_2$ [8, 13]). One should note that although THGEM detectors have shown high gas gains for single-photoelectron detection [8, 13], the latter could be considerably lower in the presence of a relatively intense radioactive background (charged particles, X-rays), due to the Raether limit. [9, 14, 15].

Recently, there has been considerable interest in THGEM operation in Ne-based gas mixtures [9, 16, 17]. The main reasons are the comparatively low operation voltages and the higher gains reached at the presence of radioactive background (higher dynamic range); the low operation voltages result in reduced discharge probability, discharge energy and charging-up effects (for a discussion on the latter, related to THGEM UV-detectors under development for COMPASS RICH upgrade, see [6, 18]).

Ne-mixtures could be naturally of general interest for a broad field of detector applications; however, the main goal of this work has been the demonstration of their applicability to RICH.

We therefore aimed at demonstrating, as a first step, that in these mixtures one could reach high photoelectron extraction from the CsI photocathode followed by their efficient collection and multiplication.

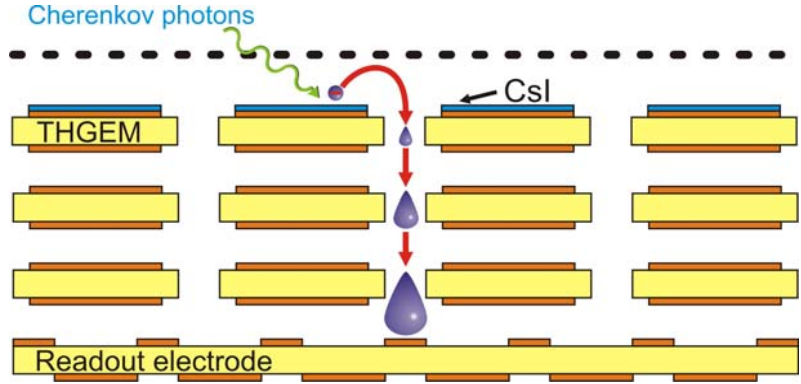

*Figure 1 - Schematic view of a THGEM-based UV-photon detector. This possible candidate for RICH, comprises a cascade of 2 or 3 THGEM electrodes, the top one coated with a reflective CsI photocathode, followed by a 2D readout anode.*

Indeed, since UV-induced photoelectrons are extracted from the CsI photocathode into gas, a fraction of them are back-scattered into the CsI, according to the gas, to the electric field and to the UV-photon energy. Further, the extracted photoelectrons drift under the electric field and a fraction of them is collected by the THGEM-dipole field into the holes. In addition to the CsI quantum efficiency (QE), the photoelectron extraction and collection efficiencies play a major role in the detector's sensitivity to single photons (through the *effective photon detection efficiency*, discussed below) and it is therefore important to maximize them, through optimization of the operation parameters.

The results of the extraction efficiency in a variety of gases can be found in [16, 19-23]. Reasonable values were reached at fields above 1kV/cm in $CH_4$, $CF_4$ and their mixtures with Ar and Ne; pure noble gases have yielded lower values. A discussion related to the backscattering in gases can be found in [16]. The collection efficiency into holes was previously measured in a THGEM with the same geometry as used in this work (see below) in several gases (Ar/$CH_4$, Ar/$CO_2$, and $CH_4$); it was found to be unity already at low THGEM gains (30, 20 and 6, respectively) [13], unlike the situation in GEMs with smaller holes [24].

It is important to note that the *absolute single-photon detection efficiency* of a detector system combines the *effective photon detection efficiency* (studied in this work) and the efficiency of counting single-photoelectron pulses above threshold. The latter very critically depends on the total gain and on the threshold of the readout electronics (system noise), in view of the typical exponential distribution of single-electron pulses. THGEM detectors have shown high charge gains both in a single- and double-step configurations [8, 13], in a variety of gases, including Ar, Xe, Ar-Xe [12], Ne [9, 17], Ar/$CH_4$, Ar/$CO_2$ [8, 13], etc.

In the present work we study the extraction and collection efficiencies of single photoelectrons from a CsI-coated THGEM electrode, in Ne-based mixtures. The results of a recent study on THGEM UV-detector operation stability under high gain, in the presence of MIPs, will be presented elsewhere [14].

## 2. Experimental Setup and Methodology

Measurements of extraction and collection efficiencies were carried out using a single THGEM (thickness of the FR4 substrate t = 0.4 mm; hole diameter d = 0.3 mm; pitch a = 0.7 mm; rim around the hole h = 0.1 mm), sandwiched between two Multi-Wire Proportional Chambers (MWPC). The MWPC had 1 mm spaced wires of 20 μm diameter. The distances between the wire plane and the cathode mesh were of 1.6 mm. The top face of the THGEM electrode was coated with a reflective CsI photocathode (200 nm thick), deposited by thermal evaporation. Measurements of extraction efficiency were carried out with the MWPC positioned above the CsI surface (MWPC$_{top}$) as shown in Figure 2. Measurements of collection efficiency of the extracted photoelectrons into the holes, were carried out with both MWPC$_{top}$ and MWPC$_{bottom}$ as shown in Figure 3.

The assembled THGEM and MWPCs were introduced into a stainless steel vessel equipped with a UV transparent Suprasil window. It was continuously flushed with 1 atm of Ne/CH$_4$ and Ne/CF$_4$ mixtures, pure CH$_4$ and pure CF$_4$. The vessel was evacuated to $10^{-5}$-$10^{-6}$ mbar with a turbo-molecular pump, prior to gas introduction. Gas composition and flow were controlled with two Mass Flow Controllers (MKS type 1179A) and a control/readout module (MKS type 247). The detector was irradiated with UV photons (185±5nm peak) from a continuously emitting ORIEL Hg(Ar) lamp. All electrodes were biased with CAEN N471A power supplies and the currents were recorded with Keithley 610 CR electrometers (current mode). In pulse-counting mode, signals were recorded with an ORTEC 124 preamplifier followed by an ORTEC 572A amplifier (shaping time = 0.5 μs) and an Amptek MCA 8000A multichannel analyser.

### 2.1 Photoelectron extraction efficiency

Figure 2 presents the experimental configuration for the extraction-efficiency ($\varepsilon_{extr}$) measurements. A positive potential was applied to the MWPC$_{top}$ interconnected meshes, establishing a drift field between the bottom mesh and the photocathode. Photoelectrons extracted from the CsI photocathode under this field were collected on the mesh. The top and bottom THGEM electrodes were interconnected and grounded through the picoamperemeter, which recorded the photocurrent, both in vacuum ($I_{vac}$) and in gas mixture ($I_{gas}$).

The extraction efficiency ($\varepsilon_{extr}$) was obtained by normalizing the current in gas ($I_{gas}$) to the vacuum current ($I_{vac}$) for each specific drift-field value:

$$\varepsilon_{extr} = \frac{I_{gas}}{I_{vac}} \tag{Eq. 1}$$

### 2.2 Single-photoelectron collection efficiency

For the more complex THGEM single-photoelectron collection efficiency ($\varepsilon_{coll}$) measurements we found convenient to follow the strategy used and discussed in detail in previous studies [13, 24, 25]. The method (Figure 3) consists on comparing, in pulse-counting mode and under identical conditions, the THGEM/MWPC$_{bottom}$ event-rate to the one measured with the MWPC$_{top}$ (known to have $\varepsilon_{coll}$ =1).

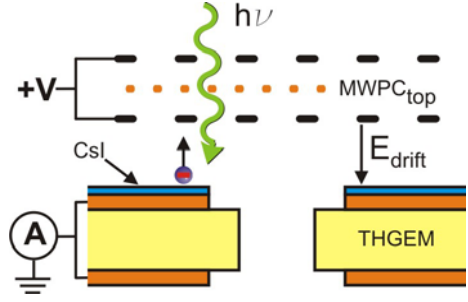

*Figure 2 - Schematic view of the experimental setup for measuring the photoelectron extraction efficiency from a CsI photocathode deposited on a THGEM.*

As shown in Figure 3, both are coupled to the same photocathode, under the same UV flux. Both, reference detector (MWPC$_{top}$) and investigated one (THGEM/MWPC$_{bottom}$) were operated at equal total gain of $\approx 10^4$, adjusted by the MWPC$_{bottom}$ voltage to have equal slopes of the total pulse-height distribution; the MWPC$_{bottom}$ role was to have the total gain sufficiently high for pulse counting even when the THGEM voltage (gain) is low. The drift field in the reference detector (Figure 3a) was adjusted to have an efficient photoelectron extraction, namely a value at the "plateau" of the extraction versus the field (see "Results"); In the investigated THGEM/MWPC$_{bottom}$ detector (Figure 3b) the drift field between the photocathode and MWPC$_{top}$ was set to zero (as discussed in [13]). The number of detected events in each configuration was evaluated by integrating the middle part of each pulse-height spectrum in order to minimize possible errors due to electronic noise contribution (lower end of the spectrum) or to secondary effects (higher end). An example of single-photoelectron pulse-height spectra in both detectors and the integration region for collection-efficiency evaluation are shown in Figure 4.

The single-electron collection efficiency of the THGEM ($\varepsilon_{coll}$) was derived from the ratio between the numbers of events measured with the THGEM/MWPC$_{bottom}$ (N$_{THGEM}$) and that in the MWPC$_{top}$ reference detector (N$_{ref}$) [13, 24, 25]:

$$\varepsilon_{coll} = \frac{N_{THGEM}}{N_{ref}}$$

(Eq. 2)

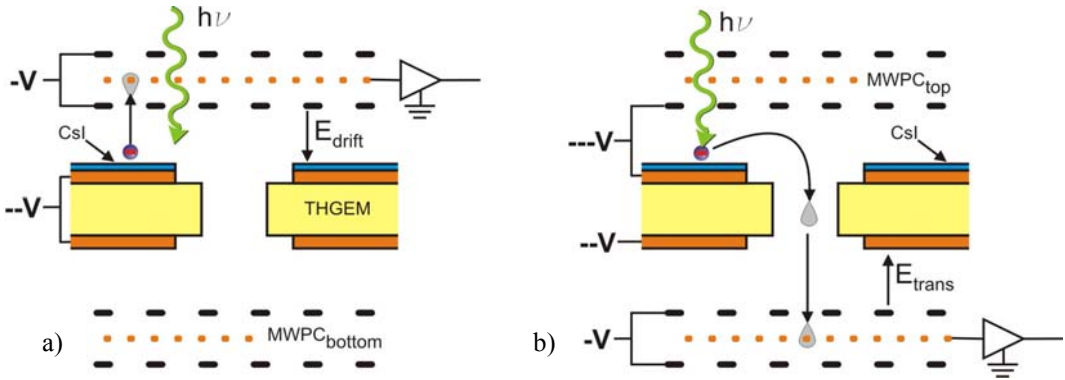

*Figure 3 - Schematic view and electrical bias of the experimental setup for measuring the single-electron collection efficiency into the THGEM holes: a) measurement of the reference pulse-height spectrum with a MWPC Reference detector; b) measurement of the pulse-height spectrum of the investigated detector.*

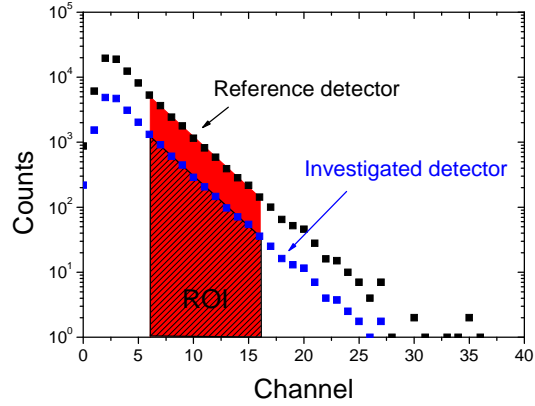

*Figure 4 - An example of single-photoelectron pulse-height distributions measured in the reference and in the investigated detectors; the respective counts, $N_{THGEM}$ and $N_{ref}$, at equal total gains (slopes) and measurement times, were derived from the integration of the ROI shown in the figure.*

## 3. Results

Figure 5 shows single-THGEM gain as function of the voltage across the electrode for atmospheric-pressure Ne/5%CH$_4$, Ne/10%CH$_4$, Ne/23%CH$_4$, Ne/5%CF$_4$ and Ne/10%CF$_4$. As general trend, the effective gain curves shift towards higher operation voltages with higher molecular-gas concentration; lower operation voltages and higher gains were reached in Ne/CF$_4$ mixtures. Note that in presence of X-ray photons or MIPs the maximum achievable gains were typically 10-fold lower [9].

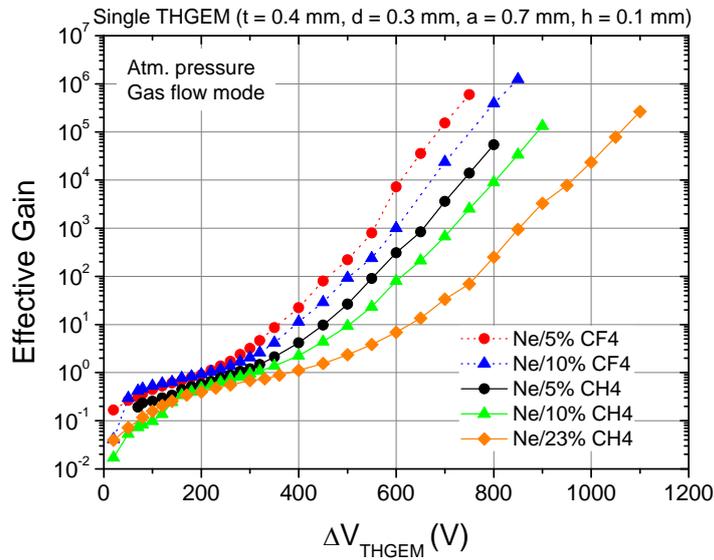

*Figure 5 - Single-THGEM gain curves, measured with UV photons in current mode, in Ne/CH$_4$ and Ne/CF$_4$ mixtures; the THGEM was coated with a CsI photocathode, irradiated with a UV lamp (185 nm peak).*

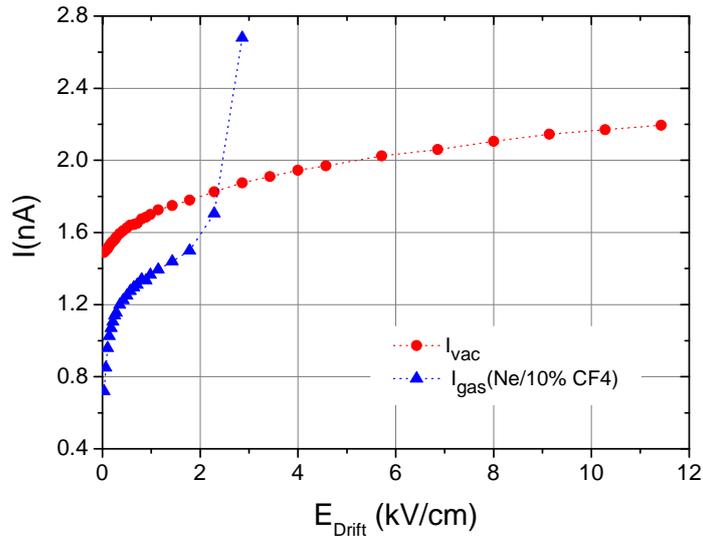

*Figure 6 - Photocathode current, in vacuum ($I_{vac}$) and gas ($I_{gas}$(Ne/10%CH₄) at 1 atm), as a function of the drift field, using the photons from a Hg(Ar) UV lamp (185 nm peak).*

Figure 6 shows typical photocurrent versus drift-field curves, in vacuum and in gas (Ne/10%CF₄ in this example), measured in the setup of Figure 2. As observed in earlier works, the vacuum photocurrent sometimes constantly increases with drift field, not reaching a real plateau; this could possibly reflect electric field penetration into the CsI bulk, enhancing extraction of photoelectrons from deeper layers. The same QE increase with the drift-field was also observed in [26, 27].

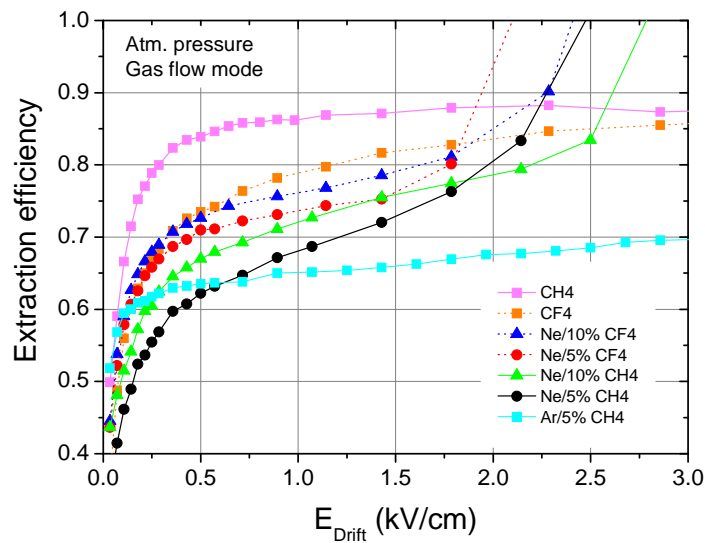

*Figure 7 - Photoelectron extraction efficiency from CsI into CH₄, CF₄, Ne/CF₄, Ne/CH₄ and Ar/CH₄ (for comparison) as function of the drift field using a UV lamp (185 nm peak). Data points measured for $\varepsilon_{extr} > 1$ are not shown.*

The photocurrent in gas is increasing with the field, up to ~2 kV/cm, due to the well known reduction of backscattering [22, 23], and then diverges due to the onset of photon feedback [22] and of charge multiplication in the gas.

The complete present data of photoelectron extraction efficiency from CsI as function of the drift field are depicted in Figure 7. The extraction efficiency in Ar/5%CH$_4$, previously investigated in THGEM and other Micropattern detectors, is shown for comparison. The efficiency "plateau" in some mixtures deviates, due to onset of secondary processes. With a reflective photocathode on the THGEM top, the field at the photocathode surface is defined by the voltage across the plate (affecting the dipole field emerging out of the holes). Examples of the field in the present electrode geometry are given in Figure 8. The CsI-coated electrode should be preferably operated at the highest possible applied voltage across the holes, though keeping in mind stability effects (charging up, micro-discharges) that are often affected at higher voltages.

Higher $\varepsilon_{extr}$ was observed in CH$_4$ compared to CF$_4$; this is in accordance with the recent independent data and simulation results of [16], but in contradiction to the data of [19]. With small additives of CH$_4$ and CF$_4$ to Ne this behaviour was reversed; i.e., $\varepsilon_{extr}$ was higher in Ne/CF$_4$ compared to Ne/CH$_4$ (Figure 7), an effect also observed and discussed in [16].

The results of single-photoelectron collection efficiency into THGEM holes are depicted in Figure 9. The photoelectron collection efficiency increases with the potential across the holes, reaching unity at THGEM gains of 25 - 1000, depending on the filling gas. Examining the data we may conclude that full collection efficiency is achieved at very realistic working conditions.

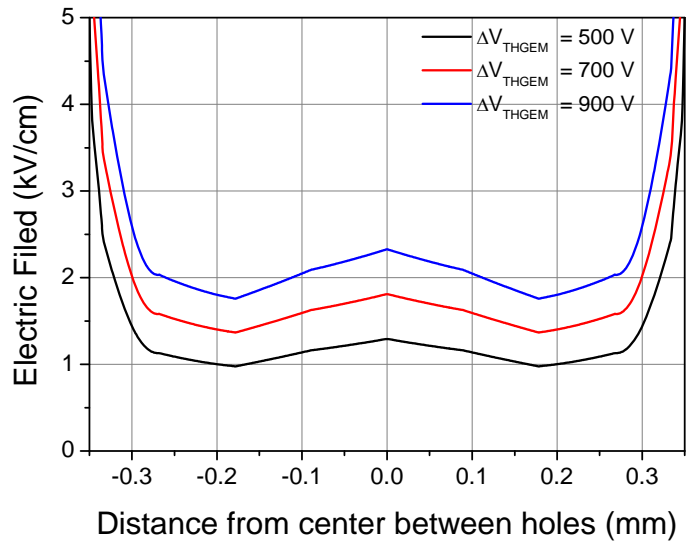

*Figure 8 - The electric field on THGEM top surface used in this work, E$_{surface}$, calculated by MAXWELL along the line interconnecting two hole centers.*

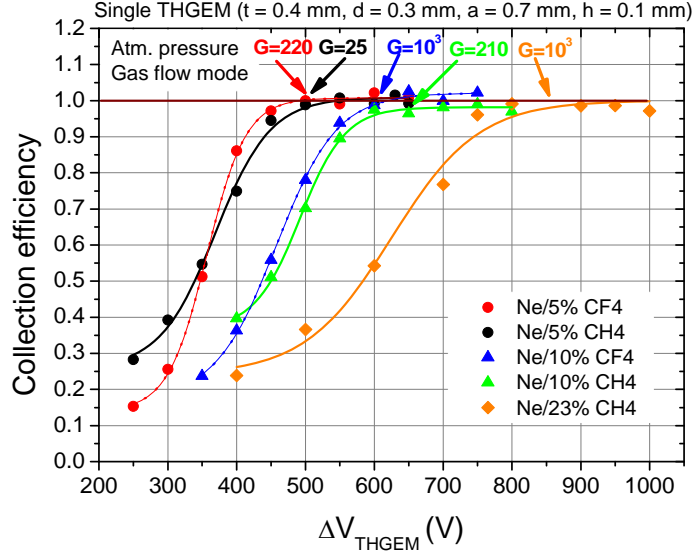

*Figure 9 - Single-photoelectron collection efficiency in Ne/CH₄ and Ne/CF₄ mixtures, measured in pulse-counting mode, versus the voltage across the THGEM; the threshold gain values for reaching full collection efficiency are indicated for each mixture.*

## 4. Discussion

The *effective photon detection efficiency* of the CsI-coated THGEM is defined as:

$$\varepsilon_{effph} = QE \times A_{eff} \times \varepsilon_{extr} \times \varepsilon_{coll} \qquad \text{(Eq. 3)}$$

where $QE$ is the CsI vacuum quantum efficiency at a given wavelength [28], $A_{\text{eff}}$ is the photocathode effective area (fraction of surface not covered by holes), $\varepsilon_{\text{extr}}$ is the extraction efficiency into gas mixture (Figure 7) and $\varepsilon_{\text{coll}}$ is the single-photoelectron collection efficiency into the THGEM[1] holes (Figure 9).

Figure 8 shows the electric field on the photocathode surface between two adjacent holes, as function of the voltage applied across the THGEM, calculated with Maxwell software [29]. As can be seen, the field at the surface exceeds 1.5 kV/cm (for maximum gains in all gases investigated; see Figure 5 and Figure 8); thus one can evaluate $\varepsilon_{\text{extr}}$ being 0.72-0.87, at 1.5 kV/cm, and larger at higher fields. (see Figure 7).

The electric field near the holes reaches values 4-5kV/cm; in this case, the electrons, drifting at the high field in the photocathode vicinity, might initiate both scintillation (photon feedback at the photocathode) and charge multiplication [9], before entering the hole. One can speculate that in case a primary electron misses the hole due to the relatively large electron diffusion in Ne mixtures, the resulting secondary electrons could compensate for that. This was demonstrated in GEM [25]. Indeed, as seen in Figure 9, the THGEM potential, and gain, requested for the onset of full collection efficiency increases with molecular-gas concentration. .

---

[1] This THGEM: t = 0.4 mm, d = 0.3 mm, a = 0.7 mm, h = 0.1 mm

It may indicate indeed that some gain outside the hole is needed to overcome the possible suppression of scintillation-induced secondary electrons by the quencher in the mixture.

Table I summarizes the main results and predictions of the effective photon detection efficiency of a THGEM-based UV-photon detector; for simplicity, the efficiency was determined for photons at wavelength of 170nm.

*Table I - Calculated Single THGEM-CsI effective photon detection efficiency at 170nm, for Ne mixtures with CF$_4$ and CH$_4$*

| *Gas* | $\Delta V_{\text{THGEM}}(V)$ | *Gain* | $QE$ 170nm | $A_{\text{eff}}$ This THGEM | $A_{\text{eff}}^2$ Optimal THGEM | $\varepsilon_{\text{extr}}$ | $\varepsilon_{\text{coll}}$ | $\varepsilon_{\text{effph}}$ This THGEM | $\varepsilon_{\text{effph}}^2$ Optimal THGEM |
|---|---|---|---|---|---|---|---|---|---|
| Ne/CH$_4$(95/5) | 800 | 5.4E4 | 0.3 | 0.54 | 0.91 | 0.73[3] | 1 | 0.12 | 0.20 |
| Ne/CH$_4$(90/10) | 900 | 1.3E5 | 0.3 | 0.54 | 0.91 | 0.79[4] | 1 | 0.13 | 0.22 |
| Ne/CF$_4$(95/5) | 750 | 6.0E5 | 0.3 | 0.54 | 0.91 | 0.76[3] | 1 | 0.12 | 0.21 |
| Ne/CF$_4$(90/10) | 850 | 1.2E6 | 0.3 | 0.54 | 0.91 | 0.83[4] | 1 | 0.14 | 0.23 |

$\varepsilon_{\text{effph}}$ values between 0.12 and 0.14 were calculated in Table I for the THGEM geometry used in this work (t = 0.4 mm, d = 0.3 mm, a = 0.7 mm, h = 0.1 mm), for a single QE value taken here as example, of 0.3 at 170nm [28]. One of the main factors in the $\varepsilon_{\text{effph}}$ value is the photocathode effective area: $A_{\text{eff}}$ = 54% in the present case. One way to increase $\varepsilon_{\text{effph}}$ is increasing $A_{\text{eff}}$, by increasing the hole pitch and decreasing the rim size ("optimal" THGEM in Table I). For example, a THGEM geometry with t = 0.4 mm, d = 0.3 mm but with a = 1mm and h = 10 μm, will have $A_{\text{eff}}$ = 91%. Based on our previous experience [13], the collection and extraction efficiencies will not change significantly with the optimal geometry, at the high-range operation voltages (gain>10$^3$). As shown in Table I, assuming that there will be no significant change in the operation voltages (indicated in the table for the current geometry), we expect for this "optimal" electrode geometry, $\varepsilon_{\text{effph}}$ values of 0.20-0.23 for the different mixtures used in this work.

It should be noted that a small rim, of 10-20 μm, was shown to result in lower attainable voltages and therefore about 10-fold lower attainable gains (this was observed in Ar/CH$_4$ and Ar/CO$_2$ mixtures [8, 18]); note that high total gains, of ~10$^6$, were recently demonstrated in a triple-THGEM with 10 microns rims, in Ar/50%CH$_4$ [6]. However, in Ne-based mixtures the operation voltages are lower and the small rim is expected to affect the maximum attainable gain in a less dramatic manner. Further studies with the "optimal" THGEM will clarify its applicability.

All results described above relate to the *effective photon detection efficiency,* $\varepsilon_{\text{effph}}$. They were obtained with a single-THGEM detector. In the case of double or triple THGEM, care must be taken of operating the first stage at the maximum possible voltage, as to keep high extraction fields. The efficiency of transferring avalanche electrons from stage to stage, does not affect $\varepsilon_{\text{effph}}$.

---

[2] Optimal THGEM: t = 0.4 mm , d = 0.3 mm , a = 1 mm; h = 0.01 mm,
[3] Value for 1.5 kV/cm photocathode electric field
[4] Values for 2kV/cm photocathode electric field

Note, that the *absolute efficiency of detecting single photons* $\varepsilon_{photon}$, e.g. in RICH, depends both on the detector's *effective photon detection efficiency* $\varepsilon_{effph}$ and on the achievable gain (detected pulses above threshold ($fp_{th}$)). In a simplified way this can be written as:

$$\varepsilon_{photon} = \varepsilon_{effph} \times fp_{th} \qquad \text{(Eq. 4)}$$

At this point we could make a very preliminary comparison between THGEM- and MWPC-based UV-photon detectors. E.g., in the MWPC of COMPASS, coated with CsI photocathode and operating with pure $CH_4$, both extraction and collection efficiencies are close to unity. The effective CsI area being 1, the *effective photon detection efficiency* being $\varepsilon_{effph} \sim$ 0.3. With a gain of 1-2 $10^4$ and electronic thresholds imposed by ionizing background and noise, $fp_{th}$ values of 0.7 at best could be reached at the experiment (0.9 at the laboratory), yielding an *absolute efficiency of detecting single photons* $\varepsilon_{photon}$ = 0.21 (0.27 at the laboratory) [30]. This may indicate, that THGEM UV-detectors of the optimal geometry (Table I), capable of operation at higher gains (> $10^5$) even at the presence of higher ionization background [14], could possibly compete with MWPC/CsI devices in future RICH applications. They are also simpler, and possibly cheaper, to produce over large surfaces. Additional studies are in course, including comparative beam tests.

## 5. Conclusions

The present work focused on the photoelectron extraction and collection efficiencies, showing that practically in all mixtures studied, extraction efficiency >72% (assuming 1.5 kV/cm surface electric field) and full collection efficiency may be reached at realistic operation conditions. *Effective photon detection efficiency* values of 12-14% at 170 nm were calculated with the present THGEM geometry, at gains >$10^5$ in the gases investigated; they are expected to increase significantly, to about 20-23%, with better optimized hole geometry (larger pitch; smaller rim size), as indicated in the discussion and in Table I.

Thus THGEM-based detectors operating in Ne-based mixtures (Ne/$CH_4$ or Ne/$CF_4$), may constitute an attractive alternative to current MWPC-based UV photon-imaging detectors for RICH applications. High gains (>$10^5$) were reached in this work in single-THGEM multipliers in Ne mixtures, at low operation voltages, with stable, discharge-free, operation - foreseeing high *absolute single-photon detection efficiencies* .

While we presented here data with single-THGEM elements, cascaded-THGEMs have the advantage of reaching higher gains at lower operating voltages per element, increasing the stability and further reducing the ion backflow [8]. The voltage on the first THGEM however, will have to be optimized as to maintain the highest possible effective single-photon detection efficiency. Large-area cascaded THGEM UV-photon detectors with reflective CsI photocathodes would have some advantages over cascaded-GEM ones [19, 31]: besides simplicity and robustness, better electron collection and transport between cascaded elements - resulting in a lower gain required per element or, alternatively, fewer cascaded elements for an equal total gain [8].

THGEM-based UV detectors are under investigations within the CERN-RD51 collaboration.

## Acknowledgments

This work was partially supported by the Israel Science Foundation, grant Nº 402/05, by the MINERVA Foundation, grant Nº8566, by the Benoziyo Center for High Energy Research and by project CERN/FP/83487/2008 under the FEDER and FCT (Lisbon). This work was pursued within the framework of the CERN RD51 collaboration. M. Cortesi warmly acknowledges the Fellowship of the Lombroso Foundation. C.D.R. Azevedo and J. Escada are supported by the FCT scholarships: SFRH / BD / 35979 / 2007 and SFRH/BD/22177/2005, respectively; both are grateful for the hospitality of the Weizmann Institute of Science group. A. Breskin is the W.P. Reuther Professor of Research in the peaceful use of Atomic Energy.